\documentclass[aps,apl,showpacs,twocolumn,superscriptaddress]{revtex4}
\usepackage{bm,color}
\usepackage{graphicx}
\usepackage{amsmath}
\pdfoutput=1

\begin{document}
\title {Writing a skyrmion on multiferroic materials}
\author{Masahito Mochizuki}
\email{mochizuki@phys.aoyama.ac.jp}
\affiliation{Department of Physics and Mathematics, Aoyama Gakuin University, Sagamihara, Kanagawa 229-8558, Japan}
\affiliation{PRESTO, Japan Science and Technology Agency, Kawaguchi, Saitama 332-0012, Japan}

\author{Yoshio Watanabe}
\affiliation{Department of Physics and Mathematics, Aoyama Gakuin University, Sagamihara, Kanagawa 229-8558, Japan}

\begin{abstract}
This paper reports on a theoretical proposal for electrical creation of magnetic skyrmions on a thin-film specimen of a multiferroic chiral magnet by local application of an electric field, instead of an electric current, via an electrode tip. This method can be traced back to the mutual coupling between skyrmion spins and the electric polarizations in multiferroics, and represents a unique technique for use in potential skyrmion-based memory devices without Joule-heating losses.
\end{abstract}
\maketitle
\begin{figure}[h]
\includegraphics[width=1.0\columnwidth]{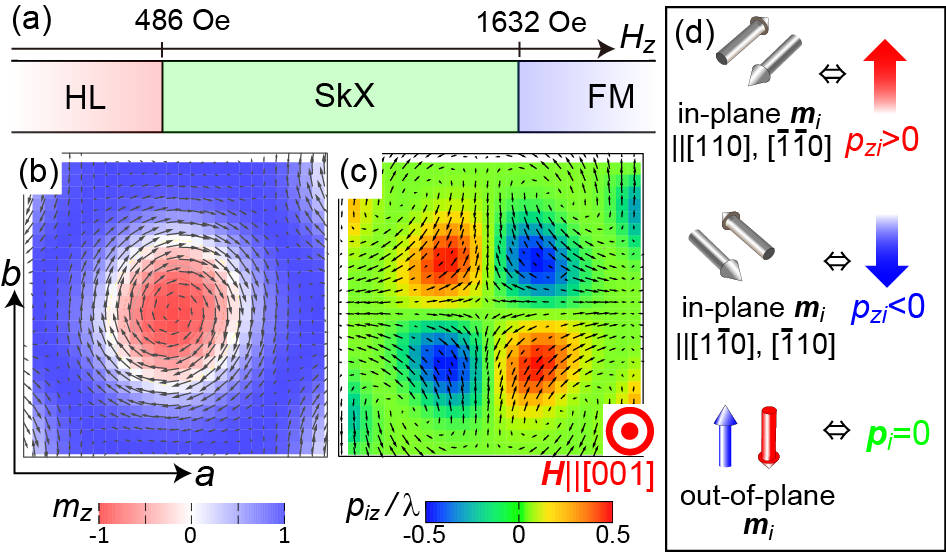}
\caption{(color online). (a) Ground-state phase diagram of the spin model~(\ref{eqn:model}) with $J$=3 meV, $D/J$=0.09, $m$=1 as a function of magnetic field $H_z$ perpendicular to the plane. HL, SkX, and FM denote helimagnetic, skyrmion-crystal, and ferromagnetic phases, respectively. (b) Magnetization configuration of a skyrmion. In-plane (out-of-plane) components of local magnetizations $\bm m_i$ are represented by arrows (colors). (c) Real-space map of local electric polarizations $\bm p_i$ induced by the skyrmion magnetizations under $\bm H$$\parallel$[001]. (d) Relationships between local $\bm m_i$ vectors and induced local $\bm p_i$ vectors.}
\label{Fig1}
\end{figure}
Magnetic skyrmions, vortex-like topological spin textures realized in chiral magnets~\cite{Bogdanov89,Muhlbauer09,YuXZ10N}, are attracting major research interest because they offer numerous advantages for applications as information carriers in future storage devices, including (1) nanoscale size, (2) topologically protected stability, (3) comparatively high transition temperatures, and (4) ultralow energy consumption when driving their motion. Specifically, it was proved that skyrmions in metals can be driven by a tiny electric current via spin-transfer torques, where the threshold current density $j_{\rm c}$ is five or six orders of magnitude lower than that required to move other magnetic structures~\cite{Jonietz10,YuXZ12,Schulz12,Everschor11,ZangJ11,Iwasaki13a,Iwasaki13b}. This finding stimulated research into novel techniques to manipulate (write, delete, read, and drive) skyrmions~\cite{Nagaosa13,Fert13,Sampaio13}.

Metallic B20 alloys have been the only class of materials that have skyrmion phase. However, an insulating skyrmion phase was discovered for the first time in 2012 in Cu$_2$OSeO$_3$~\cite{Seki12a,Adams12,Seki12b}, in which the noncollinear skyrmion spin structure induces electric polarizations through relativistic spin-orbit interactions, and the system has a multiferroic nature~\cite{Seki12a,Seki12c}. This multiferroicity offers an opportunity to manipulate skyrmions using electric fields rather than electric currents~\cite{White12,White14}, which allows us to further reduce energy consumption because electric fields in insulators do not produce Joule-heating energy losses.

To use multiferroic skyrmions as information carriers, it is necessary to establish an efficient method to create skyrmions using electric fields. In this paper, taking Cu$_2$OSeO$_3$ as an example of a skyrmion-hosting multiferroic material, we theoretically propose that skyrmions can be created very rapidly (within a few nanoseconds) on a thin film by applying a local electric field via an electrode tip. The applied electric field induces the local magnetization reversal required to form the topological spin textures by modulating the spatial configurations of the electric polarizations. This phenomenon can be traced back to magnetoelectric coupling between the swirling skyrmion spins and electric polarizations of relativistic origin. The microscopic mechanism is thus essentially distinct from the spin-transfer torque, which is the usual channel for electrical control of magnetism in {\it metallic} magnets. Our proposal may potentially lead to a unique technique for future skyrmion-based memory devices. 

The crystal structure of Cu$_2$OSeO$_3$ belongs to the chiral cubic P2$_1$3 space group, which is equivalent to B20 compounds but has different atomic coordinates~\cite{Seki12a}. There are two inequivalent Cu$^{2+}$ sites with a nominal ratio of 3:1; one is surrounded by a square pyramid of oxygen ligands, while the other is surrounded by a trigonal bipyramid. The magnetic structure of Cu$_2$OSeO$_3$ consists of a network of tetrahedra, where each tetrahedron is composed of four Cu$^{2+}$ ($S$=1/2) ions at the apexes with a three-up and one-down type ferrimagnetic spin arrangement below $T_{\rm c}$$\sim$58 K~\cite{Bos08,Belesi10}.

A continuum spin model was proposed in 1980 to describe the competition between the ferromagnetic-exchange interaction and the Dzyaloshinskii-Moriya interaction in chiral cubic magnets under an external magnetic field~\cite{Bak80}. For a numerical treatment of this continuum model, we divide the space into square meshes to obtain a classical Heisenberg model on the two-dimensional lattice~\cite{YiSD09}. The Hamiltonian of this lattice spin model is given by,
\begin{eqnarray}
\mathcal{H}_0&=&
-J \sum_{<i,j>} \bm m_i \cdot \bm m_j -g\mu_{\rm B}\mu_0 H_z \sum_i m_{iz}
\nonumber \\
& &-D \sum_{i} 
(\bm m_i \times \bm m_{i+\hat{\bm x}} \cdot \hat{\bm x} 
+\bm m_i \times \bm m_{i+\hat{\bm y}} \cdot \hat{\bm y}),
\label{eqn:model}
\end{eqnarray}
where $\bm H$=(0,0,$H_z$) is a static magnetic field perpendicular to the plane. We use $J$=3 meV and $D/J$=0.09 to reproduce the magnetic transition temperature ($\sim$60 K) and skyrmion size ($\sim$50 nm) observed in Cu$_2$OSeO$_3$~\cite{Seki12a}. All spin textures treated in this study have slow spatial variations, and their spins are thus nearly decoupled from the background lattice structure. This justifies our treatment based on a spin model with coarse-grained magnetizations on the simple square lattice although the real Cu$_2$OSeO$_3$ crystal has a rather complicated structure.

Figure~\ref{Fig1}(a) shows the ground-state phase diagram of this model as a function of the magnetic field $H_z$ normal to the thin-film plane. We find that the helimagnetic phase, the skyrmion crystal phase, and the field-polarized ferromagnetic phase emerge successively as $H_z$ increases. The critical fields here are 486 Oe and 1632 Oe, respectively. This theoretical phase diagram accurately reproduces successive phase transitions observed in thin-film samples of Cu$_2$OSeO$_3$ at low temperatures~\cite{Seki12a}.

Importantly, skyrmions in chiral magnets emerge not only in crystallized form, as observed in the skyrmion crystal phase~\cite{Muhlbauer09,YuXZ10N,Seki12a,Adams12}, but also as individual defects in the ferromagnetic background~\cite{YuXZ10N}. These skyrmion defects are also stable because of their topological protection. In the following, we demonstrate that isolated skyrmions can be created by local application of an electric field to a thin-film sample via an electrode tip.

The noncollinear skyrmion spin texture in insulators is expected to induce electric polarizations via relativistic spin-orbit interactions. The emergence of skyrmion-induced electric polarizations was indeed observed experimentally~\cite{Seki12a,Seki12c}. When local magnetization $\bm m_i = (m_{ia}, m_{ib}, m_{ic})$ is assumed for the $i$th crystallographic unit cell, the cubic crystal symmetry allows the emergence of local polarization in the following form:
\begin{eqnarray}
\bm p_i=\left(p_{ia}, p_{ib}, p_{ic} \right)
= \lambda \left(m_{ib}m_{ic}, m_{ic}m_{ia}, m_{ia}m_{ib} \right).
\label{eqn:mandp}
\end{eqnarray}
Here, the coefficient $\lambda$ is evaluated to be $\lambda$=$5.64\times10^{-27}$ $\mu$Cm using data from the polarization measurements~\cite{Seki12a,Seki12c}. Using this formula, one can calculate the spatial distributions of $\bm p_i$ induced by the skyrmion magnetizations $\bm m_i$ [Fig.~\ref{Fig1}(b)], which depends on the choice of thin-film plane. In Fig.~\ref{Fig1}(c), calculated local directions of $\bm p_i$ on the [001] plane are visualized. Comparison of Fig.~\ref{Fig1}(b) with Fig.~\ref{Fig1}(c) indicates that the in-plane $\bm m_i$ induces a finite out-of-plane $\bm p_i$, while the out-of-plane $\bm m_i$ does not induce $\bm p_i$ as summarized in Fig.~\ref{Fig1}(d). We also find that $p_{zi}>0$ when $\bm m_i$$\parallel$[110] or [$\bar{1}$$\bar{1}$0], while $p_{zi}<0$ when $\bm m_i$$\parallel$[1$\bar{1}$0] or [$\bar{1}$10].

\begin{figure*}
\includegraphics[width=2.0\columnwidth]{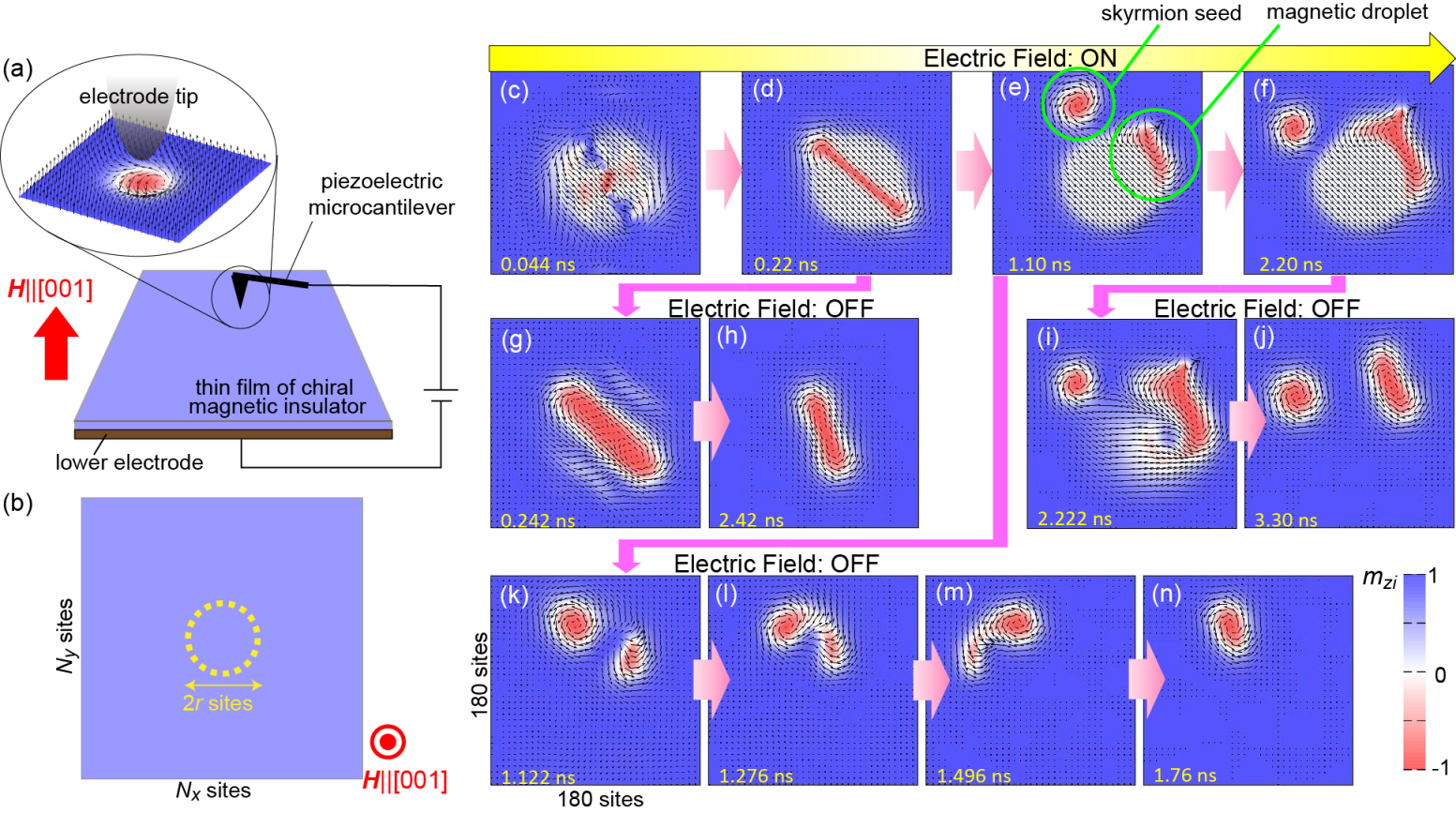}
\caption{(color online). (a) Schematic of skyrmion creation via local application of an electric field using an electrode tip. (b) System of $N$=320 $\times$ 320 sites under static magnetic field $\bm H$$\parallel$[001] with an open boundary condition that was used for the numerical simulations. An electric field $\bm E$=(0, 0, $E_z$) is applied to sites within the area indicated by the dashed circle with diameter of $2r$. (c)-(n) Simulated spatiotemporal dynamics of the magnetizations $\bm m_i$ for the electrical skyrmion creation process. The relevant area of 180 $\times$ 180 sites around the $\bm E$-field spot is magnified.}
\label{Fig2}
\end{figure*}

The coupling between magnetism and electricity offers an opportunity to manipulate magnetic skyrmions electrically by modulating the spatial distributions of their electric polarizations. We investigate dynamic processes of skyrmion creation under a local electric field [see Fig.~\ref{Fig2}(a)] by micromagnetic simulations based on the Landau-Lifshitz-Gilbert equation:
\begin{equation}
\frac{d\bm m_i}{dt}=-\bm m_i \times \bm H^{\rm eff}_i 
+\frac{\alpha_{\rm G}}{m} \bm m_i \times \frac{d\bm m_i}{dt},
\label{eq:LLGEQ}
\end{equation} 
where 
\begin{equation}
\bm H^{\rm eff}_i=-\frac
{\partial \left(\mathcal{H}_0+\mathcal{H}'(t) \right)}{\partial \bm m_i}.
\label{eq:EFFMF}
\end{equation}
Here, $\mathcal{H}_0$ is the model Hamiltonian~(\ref{eqn:model}), while $\mathcal{H}'(t)$ represents the coupling between the local polarizations $\bm p_i$ and the electric field $\bm E$, which is given by
\begin{equation}
\mathcal{H}'(t)= -\bm E(t) \cdot \sum_{i \in \mathcal{C}} \bm p_i.
\end{equation}
This equation combined with Eq.~(\ref{eqn:mandp}) describes the coupling between the electric field $\bm E$ and the magnetizations $\bm m_i$. The electric field $\bm E(t)$=(0,0,$E_z$) is applied for a fixed time to sites within a circular area $\mathcal{C}$ with diameter of $2r$ sites [see Fig.~\ref{Fig2}(b)]. Calculations are performed using a system of $N=N_x \times N_y$ sites with an open boundary condition. The Gilbert damping coefficient $\alpha_{\rm G}$ is fixed at 0.04.

Figures~\ref{Fig2}(c)-(n) show snapshots of dynamical processes of electrical skyrmion creation simulated for $H_z=1632$ Oe, $E_z=-8.52 \times 10^9$ V/m, ($N_x$, $N_y$)=(320, 320) and $2r=80$ sites. The relevant area with 180$\times$180 sites around the $\bm E$-field area is magnified. The $z$ axis is set to be parallel to $\bm H$$\parallel$[001]. Local application of the $\bm E$-field to a uniform ferromagnetic state with $m_{zi}>0$ causes local reversal of $\bm m_i$ to $m_{zi}<0$ at a small spot within the $\bm E$-field area [Fig.~\ref{Fig2}(c)]. This $\bm m_i$-reversed spot grows into a line-shaped structure [Fig.~\ref{Fig2}(d)], which has a finite skyrmion number $Q=-1$. Therefore, if we switch off the $\bm E$-field at this point, we obtain a single skyrmion after relaxation of the spatial distribution of $\bm m_i$, as shown in the process of Figs.~\ref{Fig2}(c)$\rightarrow$(d)$\rightarrow$(g)$\rightarrow$(h).

However, if we continue to apply the $\bm E$-field, this $\bm m_i$-reversed line becomes kicked out of the $\bm E$-field area because the area of $m_z=-1$ with $p_z$$\sim$0 is energetically unfavorable under the condition of $E_z<0$, and breaks into two parts; the first is a skyrmion seed with perfectly reversed $\bm m_i$, while the other is a magnetic droplet with imperfect $\bm m_i$ reversal~\cite{Mohseni13}. The skyrmion number is $Q=-1$ for the skyrmion seed, but it is zero for the magnetic droplet. The total skyrmion number is thus conserved upon collapse of the $\bm m_i$-reversed line. If we switch off the $\bm E$-field at this point, the skyrmion seed grows into a skyrmion, whereas the magnetic droplet vanishes or is absorbed by the skyrmion seed. Eventually, we obtain a single skyrmion again, as shown in the process of Figs.~\ref{Fig2}(c)$\rightarrow$(d)$\rightarrow$(e)$\rightarrow$(k)$\rightarrow$(l)$\rightarrow$(m)$\rightarrow$(n). If we continue to apply the $\bm E$-field, the magnetic droplet then grows into a skyrmion seed with perfect $\bm m_i$ reversal [Fig.~\ref{Fig2}(f)]. In this case, we obtain a pair of skyrmions after the $\bm E$-field is switched off, as shown in the process of Figs.~\ref{Fig2}(c)$\rightarrow$(d)$\rightarrow$(e)$\rightarrow$(f)$\rightarrow$(i)$\rightarrow$(j). We can therefore control the number of skyrmions created by tuning the duration of the $\bm E$-field application.

\begin{figure*}
\includegraphics[width=2.0\columnwidth]{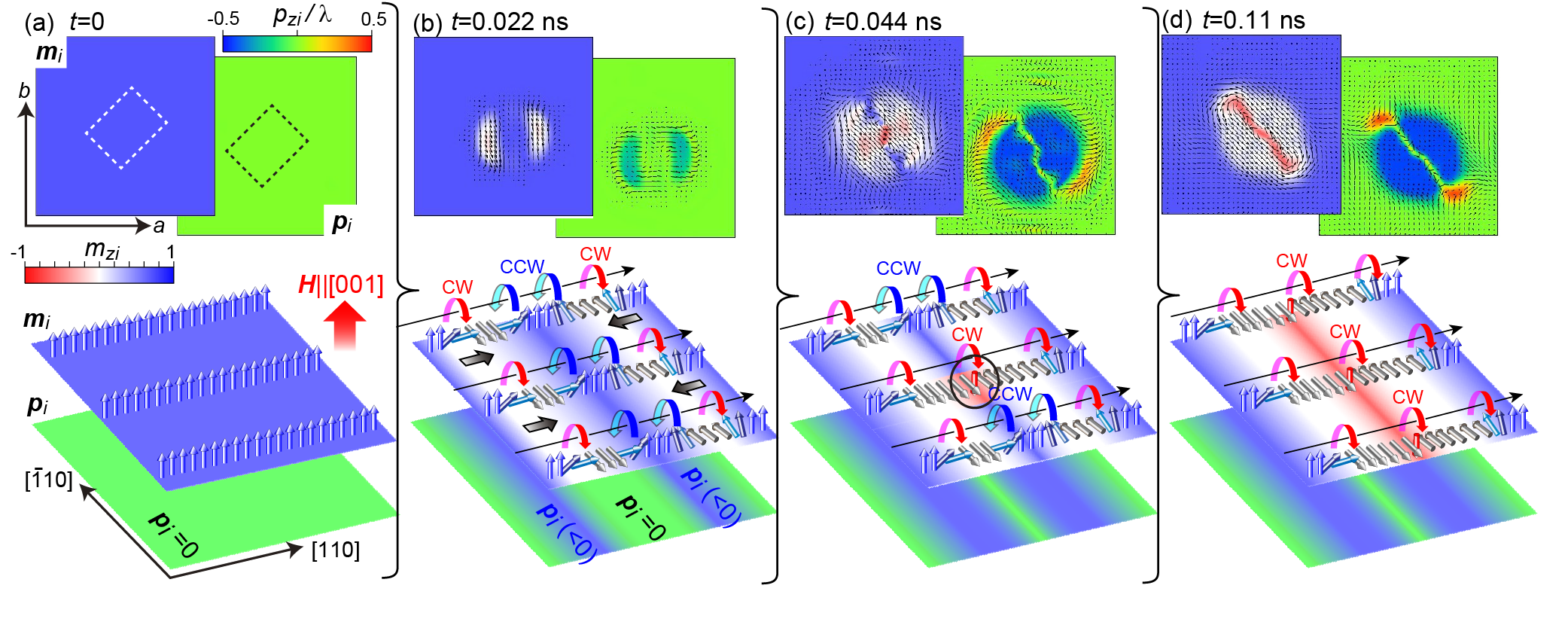}
\caption{(color online). Dynamical process of the $\bm E$-field-induced local magnetization reversal. Snapshots of simulated spatial distributions of the $\bm m_i$ and $\bm p_i$ vectors are shown in the upper panels, while schematics of their alignments within the relevant area are shown in the lower panels at selected times: (a) $t$=0, (b) $t$=0.022 ns, (c) $t$=0.044 ns, and (d) $t$=0.11 ns. The spatial spot at which nucleation of the $\bm m_i$ reversal occurs is indicated by a solid circle in the lower panel of (c).}
\label{Fig3}
\end{figure*}
\begin{figure*}
\includegraphics[width=2.0\columnwidth]{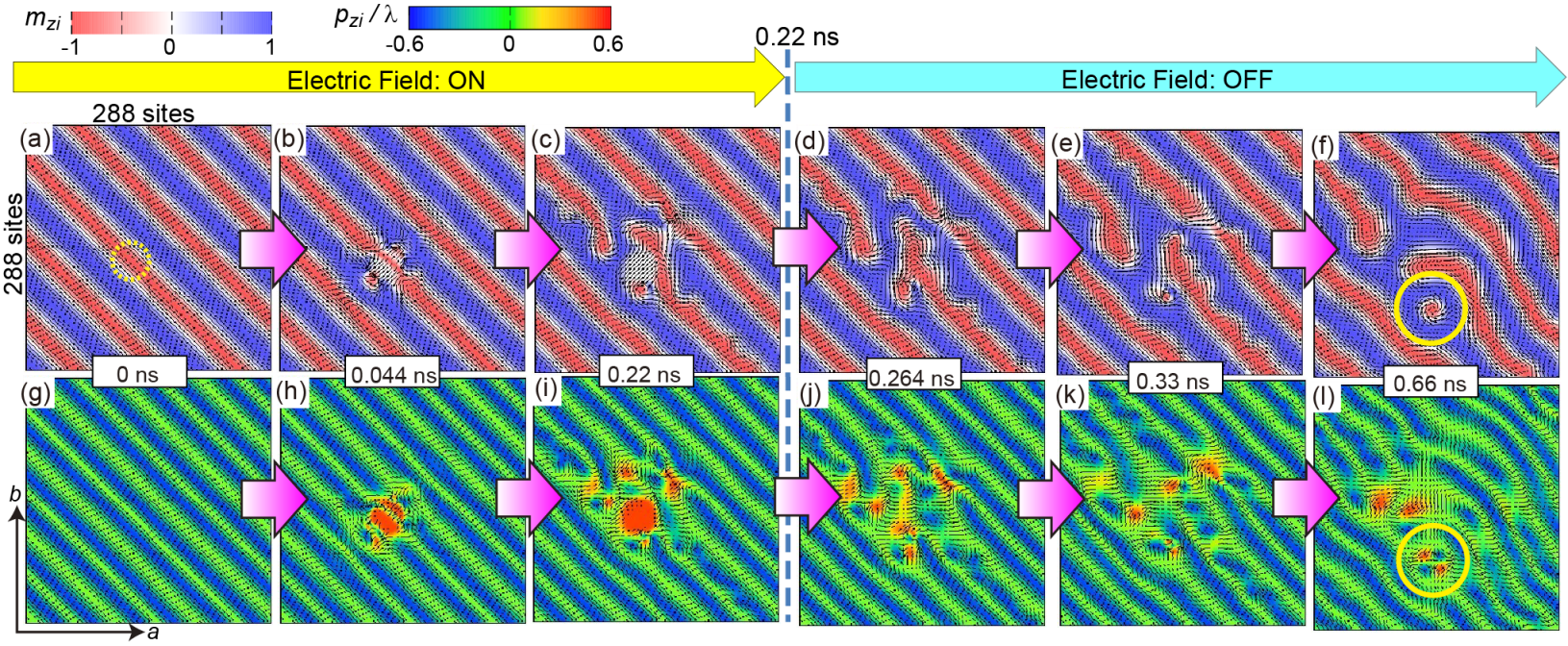}
\caption{(color online). Snapshots of simulated spatiotemporal dynamics of magnetizations $\bm m_i$ (upper panels) and polarizations $\bm p_i$ (lower panels) for electrical skyrmion creation on a helimagnetically-ordered thin film. The $\bm E$ field is applied locally to sites within the circular area indicated by the dashed circle in (a), with diameter of $2r$=40 sites.}
\label{Fig4}
\end{figure*}
We find that the local reversal of $\bm m_i$ is a key element in the creation of the topological skyrmion texture. Figures~\ref{Fig3}(a)-(d) show the $\bm m_i$ reversal process induced by the $\bm E$-field; the two upper panels are simulated snapshots of the $\bm m_i$ and $\bm p_i$ distributions, while the two lower panels are schematics of the $\bm m_i$ and $\bm p_i$ alignments in the relevant areas indicated by the dashed lines in Fig.~\ref{Fig3}(a). Initially, the $\bm m_i$ vectors are uniformly oriented in the $\bm H$$\parallel$[001] direction, and the polarizations $\bm p_i$ are thus uniformly zero [see Fig.~\ref{Fig3}(a)]. When we apply a negative $\bm E$-field with $E_z<0$, the $\bm m_i$ vectors near the periphery of the $\bm E$-field area begin to rotate to point in the in-plane directions ([$\bar{1}$10] or [1$\bar{1}$0] direction) [see Fig.~\ref{Fig3}(b)], because the $p_{zi}<0$ induced by the in-plane-oriented $\bm m_i$ are favorable when $E_z<0$. This $\bm m_i$-rotation occurs in the clockwise (CW) direction around the periphery upon propagation in the [110] direction, indicated by solid arrows, because it is energetically favored by the Dzyaloshinskii-Moriya interactions of Eq.~(\ref{eqn:model}) when $D>0$. This type of $\bm m_i$ reorientation, however, necessarily causes rotating $\bm m_i$ alignments in the counterclockwise (CCW) direction inside the $\bm E$-field area [see Fig.~\ref{Fig3}(b)], which are energetically unfavorable for the Dzyaloshinskii-Moriya interactions. To resolve this energy cost, the $\bm m_i$ at a specfic site, indicated by the solid circle in Fig.~\ref{Fig3}(c), flops to be reversed and thus  make the rotation sense clockwise. This nucleated $\bm m_i$-reversed spot immediately grows into a line-shaped area as shown in Fig.~\ref{Fig3}(d).

Our simulations demonstrate that an isolated skyrmion can also be created in the helimagnetic state. Figures~\ref{Fig4}(a)-(l) show snapshots of the $\bm m_i$ and $\bm p_i$ distributions for the simulated dynamical process of electrical skyrmion creation in the helimagnetic state with propagation vector $\bm q$$\parallel$[110]. In this case, we set $H_z$=486 Oe, $E_z= 1.2 \times 10^{10}$ V/m, ($N_x$, $N_y$)=(288, 288), and $2r=40$ sites. We find that the sign of $E_z$ must be chosen appropriately, depending on the $\bm q$ vector, to create a skyrmion in the helimagnetic state. The helical $\bm q$ vector in a chiral cubic magnet like Cu$_2$OSeO$_3$ tends to be oriented in the [110] or [$\bar{1}$10] direction on the [001] plane because of the higher-order magnetic anisotropy. We also note that the sign of $E_z$ determines the sign of $p_{zi}$ and subsequently the direction of the in-plane-oriented $\bm m_i$. A skyrmion can be created when the in-plane-oriented $\bm m_i$ in the $\bm E$-field area is perpendicular to the in-plane components of the helically ordered $\bm m_i$, or is equivalently parallel to the $\bm q$ vector, which efficiently cuts the helimagnetic alignment of $\bm m_i$ to create a skyrmion seed. $E_z$ should therefore be positive (negative) when $\bm q$$\parallel$[110] ($\bm q$$\parallel$[$\bar{1}$10]).

In summary, we have theoretically demonstrated that by using magnetoelectric coupling in multiferroic chiral magnets, isolated magnetic skyrmions can be created electrically through local electric-field application via an electrode tip on ferromagnetically- or helimagnetically-ordered thin-film samples under $\bm H$$\parallel$[001]. To date the creation of skyrmions in spin-polarized scanning tunneling microscopy has been successfully demonstrated on a metallic sample~\cite{Romming13}, which indicates that the experimental technique required to realize our proposed method is already established. The proposed method, which is based on an electric field in an insulator, is free from Joule-heating losses and is thus expected to be useful for future spintronics applications of multiferroic skyrmions in high-efficiency magnetic storage devices. While the simulations presented here were all performed at $T$=0, the proposed mechanism is expected to work at finite temperatures below $T_{\rm c}$. Also, thermal fluctuations may reduce the threshold $\bm E$-field strength for $\bm m_i$-reversal by reducing the free-energy barriers. However, $T_{\rm c}$ of $\sim$58 K for Cu$_2$OSeO$_3$ is quite low for real device applications, and further research to identify suitable materials with higher $T_{\rm c}$ values is needed.

The authors would like to thank A. Rosch and N. Romming for enlightening discussions. This research was in part supported by JSPS KAKENHI (Grant Nos. 25870169 and 25287088).

\end{document}